\documentclass[aps,pre,amsmath,amssymb,showpacs,twocolumn,draft]{revtex4}

\usepackage{bm}

\begin{document}

\newcommand{\pp}[1]{\phantom{#1}}
\newcommand{\be}{\begin{eqnarray}}
\newcommand{\ee}{\end{eqnarray}}
\newcommand{\ve}{\varepsilon}
\newcommand{\vs}{\varsigma}
\newcommand{\Tr}{{\,\rm Tr\,}}
\newcommand{\pol}{\frac{1}{2}}
\newcommand{\ba}{\begin{array}}
\newcommand{\ea}{\end{array}}
\newcommand{\bear}{\begin{eqnarray}}
\newcommand{\eear}{\end{eqnarray}}
\title{
Entangled-state cryptographic protocol that remains secure even if nonlocal hidden variables exist and can be measured with arbitrary precision
}
\author{Diederik Aerts $^1$, Marek Czachor $^{1,2}$, and Marcin Paw{\l}owski $^{2}$\\
$^1$ Centrum Leo Apostel (CLEA) and Foundations of the Exact Sciences (FUND)\\
Vrije Universiteit Brussel, 1050 Brussels, Belgium\\
$^2$ Katedra Fizyki Teoretycznej i Metod Matematycznych\\
Politechnika Gda\'nska, 80-952 Gda\'nsk, Poland}

\begin{abstract}
Standard quantum cryptographic protocols are not secure if one assumes that nonlocal hidden variables exist and can be measured with arbitrary precision. The security can be restored if one of the communicating parties randomly switches between two standard protocols.
\end{abstract}

\maketitle

It is known that quantum mechanics can be without difficulty replaced by a nonlocal hidden-variable theory
\cite{Bohm1,BohmHiley,Bohm2,DHKV88,Holland,Valentini,Durt,Duerr,f1,f2,f5,f6,f7}. Simultaneously, it is a rather popular belief that an exact knowledge of nonlocal hidden variables would destroy security of quantum cryptography. In this note we do not want to get into the crossfire of the discussion if such exact knowledge is possible or not in a hidden-variable theory that is exactly equivalent to standard quantum mechanics. Perhaps, the problem we discuss is present only in theories that are `infinitesimally close' to quantum mechanics. We are not experts in nonlocal hidden variables and, keeping in mind that impossibility proofs may only prove our lack of imagination, prefer to assume the worst possible scenario: Nonlocal hidden variables exist and can be exactly known to our enemies. We harness the nonlocality as a means of protection by a simple modification of a quantum protocol. The idea is illustrated on nonrelativistic Bohm theory, but one can argue that the effect is typical of all nonlocal hidden-variable theories. 

Bohm's theory in its simplest nonrelativistic version \cite{Bohm1} involves nonlocal hidden variables $\bm q_j(\bm x_1,\dots,\bm x_n,t)$ that have a meaning of trajectories. 
The Schr\"odinger equation for an $n$-particle wave function $\psi(\bm x_1,\dots,\bm x_n,t)$
is related by the rule $\psi=R\exp(iS/\hbar)$ to the system of partial differential equations involving Hamilton-Jacobi and continuity equations 
\be
\textstyle
\partial S/\partial t
+
\sum_{j=1}^n m_j\bm v_j^2/2+Q+V
&=&0,\\
\textstyle
\partial \rho/\partial t
+
\sum_{j=1}^n \bm\nabla_j(\rho \bm v_j)
&=&
0.
\ee
$\rho=R^2$ is the density of particles, $\bm v_j=\bm\nabla_jS/m_j$ the velocity if a $j$-th particle,  $V=V(\bm x_1,\dots,\bm x_n,t)$  the usual potential, and 
$Q=-\hbar^2\sum_{j=1}^n \bm\nabla_j^2 R/(2m_jR)$ is the so-called quantum potential. 
The hidden trajectories are found by integrating the `guidance equation' 
$d\bm q_j/dt=\bm v_j$. 
If the particles are not entangled (and thus not interacting via $V$), that is the wave fuction takes the product form
$\psi(\bm x_1,\dots,\bm x_n,t)=\psi_1(\bm x_1,t)\dots\psi_n(\bm x_n,t)$, then 
$Q=\sum_{j=1}^n Q_j$ where $Q_j=-\hbar^2\bm\nabla_j^2 R_j/(2m_jR_j)$. Such particles cannot communicate via the quantum potential. However, for entangled states the particles do interact via $Q$ even if in the sense of $V$ they are uninteracting. Systems described by entangled states are thus nonlocal: The dynamics of a $k$-th particle depends on what happens to the remaining $n-1$ particles. What is important, the influences remain within the entangled system. The quantum potential is a useful conceptual tool in this context, but the Bohm theory  needs only the Schr\"odinger and guidance equations.

An eavesdropper (Eve) attempting to read the secret code via the quantum potential would have to  get entangled (in the quantum sense) with the information channel and would be detected by the usual means, say, an Ekert-type procedure \cite{E91,BBM92}. If the eavesdropper does not get entangled, the quantum potential will not carry the information she needs.

Let us now assume that Eve can know the hidden trajectory $\bm q(t)$ of the particle carrying the key between the two communicating parties. A Bohmian analysis of spin-1/2 measurements performed via Stern-Gerlach devices \cite{DHKV88,Holland} shows that the knowledge of 
$\bm q(t_0)$ at some initial time $t_0$ {\it uniquely\/} determines the results of future measurements of spin in any direction (\cite{Holland}, pp. 412-415). 
The single-particle schemes of the BB84 variety \cite{BB84} are thus clearly insecure from this perspective. To make matters worse, a similar statement can be deduced from the analysis of two-electron singlet states described in detail in Chapter 11 of \cite{Holland}. If two Stern-Gerlach devices are aligned along the same direction $(0,0,1)$ and the particles propagate toward the Stern-Gerlach devices of Alice and Bob with velocities 
$\bm v_1=(0,-|\bm v_1|,0)$ and 
$\bm v_2=(0,|\bm v_2|,0)$, respectively, then the results of spin measurements are always opposite (that is why we use them for generating the key) but are uniquely determined by the sign of $z_1(t_0)-z_2(t_0)$, where the respective trajectories are 
$\bm q_1(t)=(0, y_1(t),z_1(t))$ and 
$\bm q_2(t)=(0, y_2(t),z_2(t))$ (cf. the discussion on p. 470 in \cite{Holland}). The result agrees with the analysis of \cite{Valentini}. 

Still, if one looks more closely at the derivation given in \cite{Holland} one notices that the two particles interact with {\it identical\/} magnetic fields. We can weaken this assumption. Following \cite{Holland} we assume that the time of interaction with the Stern-Gerlach magnets is $T$, the particles are identical, their magnetic moments and masses 
equal $\mu$ and $m$, and the initial wave functions are Gaussians of half-width $\sigma_0$ in the $z$ directions. We also assume that Alice's Stern-Gerlach produces the field $\bm B_1(\bm q_1)=(0,0,B_0+B z_1)$ but, contrary to \cite{Holland}, the Bob field is taken as 
$\bm B_2(\bm q_2)=
\kappa(0,0,B_0+B z_2)$, where $\kappa$ is a real number (in \cite{Holland} $\kappa=1$). 
Then the velocities in the $z$ direction $(0,0,1)$ read 
\begin{widetext}
\be \label{7}
dz_1(t)/dt
&=&\hbar^2 t\, z_1(t)/\big(4 m^2 \sigma_0^4 \ve(t)\big) +\big(m \ve(t)\big)^{-1}B\mu T
\tanh \big[\big(m\sigma_0^2 \ve(t)\big)^{-1}\big(z_1(t)-\kappa z_2(t)\big)B \mu T  t\big], 
\\ \label{8}
dz_2(t)/dt
&=&
\hbar^2 t\, z_2(t)/\big(4 m^2 \sigma_0^4 \ve(t)\big) - \big(m \ve(t)\big)^{-1}\kappa B\mu T 
\tanh \big[\big(m\sigma_0^2 \ve(t)\big)^{-1}\big(z_1(t)-\kappa z_2(t)\big)B \mu T  t\big],
\ee
\end{widetext}
where
$\ve(t)=1+\frac{\hbar^2 t^2}{4\sigma_0^4m^2}$. The above formulas differ from Eqs. (11.12.15), 
(11.12.16) found in \cite{Holland} only by the presence of $\kappa$. This apparently innocent generalization has a fundamental meaning for the quantum protocol. For reasons that are identical to those discussed by Holland in his book
the signs of spin found in the labs of Alice and Bob depend on the sign of the term under tanh. However, as opposed to the case of identical magnetic fields this sign is controlled not 
only by the initial values of $z_1(t_0)$ and $z_2(t_0)$, in principle known to Eve, but also by the parameter $\kappa$ which is known only to Bob. If $|\kappa|\gg 1$ then the sign of this term is practically controlled by the sign of $\kappa$ (recall that the range of $z_1$ is limited by the size of the Gaussian). Choosing the sign of $\kappa$ randomly, Bob can flip the spin of the particle which is already in the lab of Alice and is beyond the control of Eve. Eve knows, by looking at $z_1(t_0)$ and 
$z_2(t_0)$, what will be the result of Alice's measurement if sign$(\kappa)=+1$, and that if sign$(\kappa)=-1$ the result will be opposite. But she {\it does not know\/} this sign if Bob keeps it secret! It follows that she gains nothing by watching the trajectory. But Bob always knows the result of Alice's measurement due to the EPR correlations. If he keeps $\kappa>0$ then Alice got the result opposite to what he found in his lab because $\bm B_1$ and 
$\bm B_2$ are parallel; if he takes  $\kappa<0$ then both Alice and Bob find the same number because $\bm B_1$ and $\bm B_2$ are anti-parallel.
And this is sufficient for producing the key. 

Let us finally clarify here one point that can be easily misunderstood at a first reading of our protocol. 
In the Ekert protocol we have 
four settings of experimental devices that are used for testing the Bell inequality: 
$(A,B)$, $(A,B')$, $(A',B)$, $(A',B')$. This part of the data cannot be used for 
producing the key. We need one more setting, say $(C,C)$, that will be used for the key. 
In our protocol we have in addition the setting $(C,-C)$. 
One can even think of our protocol as a version of the Ekert one but with two alternative measurements corresponding to the {\it same\/} observable. 

What is important, from the 
hidden-variable point of view we {\it can\/} predict what will be the results (for each pair 
of particles) of $(C,C)$ and $(C,-C)$ measurements. 
If the initial hidden variables are such that the results of the 
measurement of $(C,C)$ would yield, say, $(C,C)= (+,-)$ then a result of $(C,-C)$ is not 
$(C,-C)= (+,+)$, as one might naively expect,  but 
$(C,-C)= (-,-)$. It is the bit of Bob that does not change even though it is Bob who flips his device! This is how the nonlocality works and why Eve does not know the key.

The work of MC and MP is a part of the Polish Ministry of Scientific Research and Information Technology (solicited) project PZB-MIN-008/P03/2003. We acknowledge the support  of the Flemish Fund for Scientific Research (FWO Project No. G.0335.02). We are indebted to S. Goldstein, P. R. Holland, R. Tumulka, P. Horodecki, D. Mayers, and the referees for their comments on preliminary versions of this work.


\begin{references}
\bibitem{Bohm1}D. Bohm, Phys. Rev. {\bf 85}, 166 (1952).
\bibitem{BohmHiley}D. Bohm and B. J. Hiley, {\it The Undivided Universe\/}, Routledge, London, 1993.
\bibitem{Bohm2}D. Bohm, Phys. Rev. {\bf 89}, 458 (1953).
\bibitem{DHKV88}C. Dewdney, P. R. Holland, A. Kyprianidis, and J.-P. Vigier, Nature {\bf 336}, 536-44 (1988).
\bibitem{Holland}P. R. Holland, {\it The Quantum Theory of Motion\/} (Cambridge, Cambridge University Press, 1993).
\bibitem{Valentini}A. Valentini, Pramana - J. Phys. {\bf 59}, 269 (2002).
\bibitem{Durt}T. Durt and Y. Pierseaux, Phys. Rev. A {\bf 66}, 052109 (2002).
\bibitem{Duerr}D. D\"urr, S. Goldstein, and N. Zanghi, J. Stat. Phys. {\bf 67}, 843 (1992); 
J. Stat. Phys. {\bf 68}, 259 (1993)
\bibitem{f1}D. D\"urr, S. Goldstein, R. Tumulka, and N. Zanghi, Phys. Rev. Lett. {\bf 93}, 090402 
(2004).
\bibitem{f2}G. Horton and  C. Dewdney, J. Phys. A: Math. Gen. {\bf 37}, 11935 (2004).
\bibitem{f5}P. Holland and  C. Philippidis, Phys. Rev. A {\bf 67}, 062105 (2003).
\bibitem{f6}G. D. Barbosa and  N. Pinto-Neto, Phys. Rev. D {\bf 69}, 065014 (2004).
\bibitem{f7}P. Holland, Ann. Phys. (NY) {\bf 315}, 503 (2005).
\bibitem{E91}A. K. Ekert, Phys. Rev. Lett. {\bf 67}, 661 (1991).
\bibitem{BBM92}C. H. Bennett, G. Brassard, and N. D. Mermin, Phys. Rev. Lett. {\bf 68}, 557 (1992). 
\bibitem{BB84}C. H. Bennett and G. Brassard, Proc. IEEE Int. Conf. on Computers, Systems, and Signal Processing, Bangalore, India (New York, IEEE, 1984).
\end{references}
\end{document}